\documentclass[useAMS,usenatbib]{mn2e}
\usepackage{graphicx}
\def \th {\thinspace}

\def \degmark {^\circ}

\title[Jets in Cyg\th X-2] 
{Radio and X-ray observations of jet ejection in Cygnus\th X-2}
\author[Spencer et al.] 
{R. E. Spencer$^{1}$, A. P. Rushton$^{2}$, M. Ba\l uci\'nska-Church$^{3}$, Z. Paragi$^{4}$, 
\and N. S. Schulz$^{5}$, J. Wilms$^{6}$, G. G. Pooley$^{7}$ and M. J. Church$^{3}$\\
$^{1}$ Jodrell Bank Observatory, University of Manchester, Macclesfield, Cheshire, SK11\th 9DL\\
$^{2}$ School of Physics \& Astronomy, University of Southampton, Highfield, SO17\th Southampton\\
$^{3}$ School of Physics \& Astronomy, University of Birmingham, Birmingham B15\th 2TT\\
$^{4}$ Joint Institute for VLBI in Europe, Postbus 2, NL-7990\th AA, Dwingeloo, The Netherlands\\
$^{5}$ Kavli Institute for Astrophysics and Space Research, Massachusetts Institute of Technology, Cambridge MA\th 02139, USA\\
$^{6}$ Dr Karl Remeis-Sternwarte, Astronomisches Institut der Universit\"at Erlangen-N\"urnberg, Sternwartestrasse 7, 96049 Bamberg, Germany\\
$^{7}$ Cavendish Laboratory, Cambridge University, J J Thomson Avenue, Cambridge CB3\th 0HE.}
\begin{document}

\date{Accepted 2013 July 10. Received 2013 July 01; in original form 2013 June 03}

\pagerange{\pageref{firstpage}--\pageref{lastpage}} \pubyear{2013}

\maketitle

\label{firstpage}

\begin{abstract}
The ejection of a relativistic jet has been observed in the luminous Galactic low mass X-ray binary 
Cygnus\th X-2. Using high resolution radio observations, a directly resolved ejection event 
has been discovered while the source was on the Horizontal Branch of the Z-track. Contemporaneous 
radio and X-ray observations were made with the European {\it VLBI} Network at 6 cm and the 
{\it Swift} X-ray observatory in the 0.3 -- 10 keV band. This has been difficult to achieve
because of the previous inability to predict jet formation. Two sets of $\sim$10 hr observations
were spaced 12 hr apart, the jet apparently switching on during Day 1. The radio results
show an unresolved core evolving into an extended jet. A preliminary value of jet velocity v/c of 0.33
$\pm 0.12$ was obtained, consistent with previous determinations in Galactic sources. Simultaneous radio 
and X-ray lightcurves are presented and the X-ray hardness ratio shows the source to be on 
the Horizontal Branch where jets are expected. The observations support our proposal that jet 
formation can in future be predicted based on X-ray intensity increases beyond a critical value. 
\end{abstract}


\section{Introduction}

Cygnus\th X-2 is a bright Galactic low mass X-ray binary belonging to the Z-track
group of sources having X-ray luminosities at or above the Eddington
limit (Schulz, Hasinger \& Tr\"umper 1989; Hasinger \& van der Klis 1989). 
These sources display three tracks in hardness-intensity: the Horizontal
Branch (HB), Normal (NB) and Flaring Branch (FB) showing that major physical 
changes which have not been understood take place within the sources. Radio jets have
been observed from these sources, notably in Sco\th X-1 (Fomalont et al. 2001) 
but predominantly when the source is on the HB (Penninx 1989). There is also strong interest
in the disk-jet connection (e.g. Migliari et al. 2011).

Based on an emission model for low mass X-ray binaries embodying the extended nature of
the accretion disk corona (Church \& Ba\l uci\'nska-Church 2004), we have proposed a 
detailed and novel physical explanation of the Z-track sources
(Church et al. 2012). On the NB and HB, there is a strong increase of neutron star temperature
and so of radiation pressure which disrupts the inner disk, launching the jets. Based on this
improved understanding it becomes important to obtain simultaneous radio and X-ray
observations which allow detailed investigation of changes at the inner accretion disk
at the time of jet launching and subsequent to this launch. The present observations 
then provide an adequate strategy to observe these events. It is moreover important 
to investigate differences between neutron star and black hole systems
as the latter do not have a stellar surface.

We observed Cygnus\th X-2 during two 10 hr periods separated by 12 hr in 2013 February 2013,
in both radio and X-rays. The observing strategy was designed to maximise the
likelihood of observing Z-track motion, and to capture the location of the source on the HB.

\section{The Radio Observations and Results}

Radio observations of Cygnus\th X-2 were made at a wavelength of 6 cm with the
European VLBI Network (EVN) on 2013 February 22$^{\rm nd}$ and 23$^{\rm rd}$, using the electronic
VLBI technique (e-VLBI, see e.g. Rushton et al. 2007) where real-time 
\begin{figure*}                                                         
\begin{center}
\includegraphics[width=75mm,height=55.7mm]{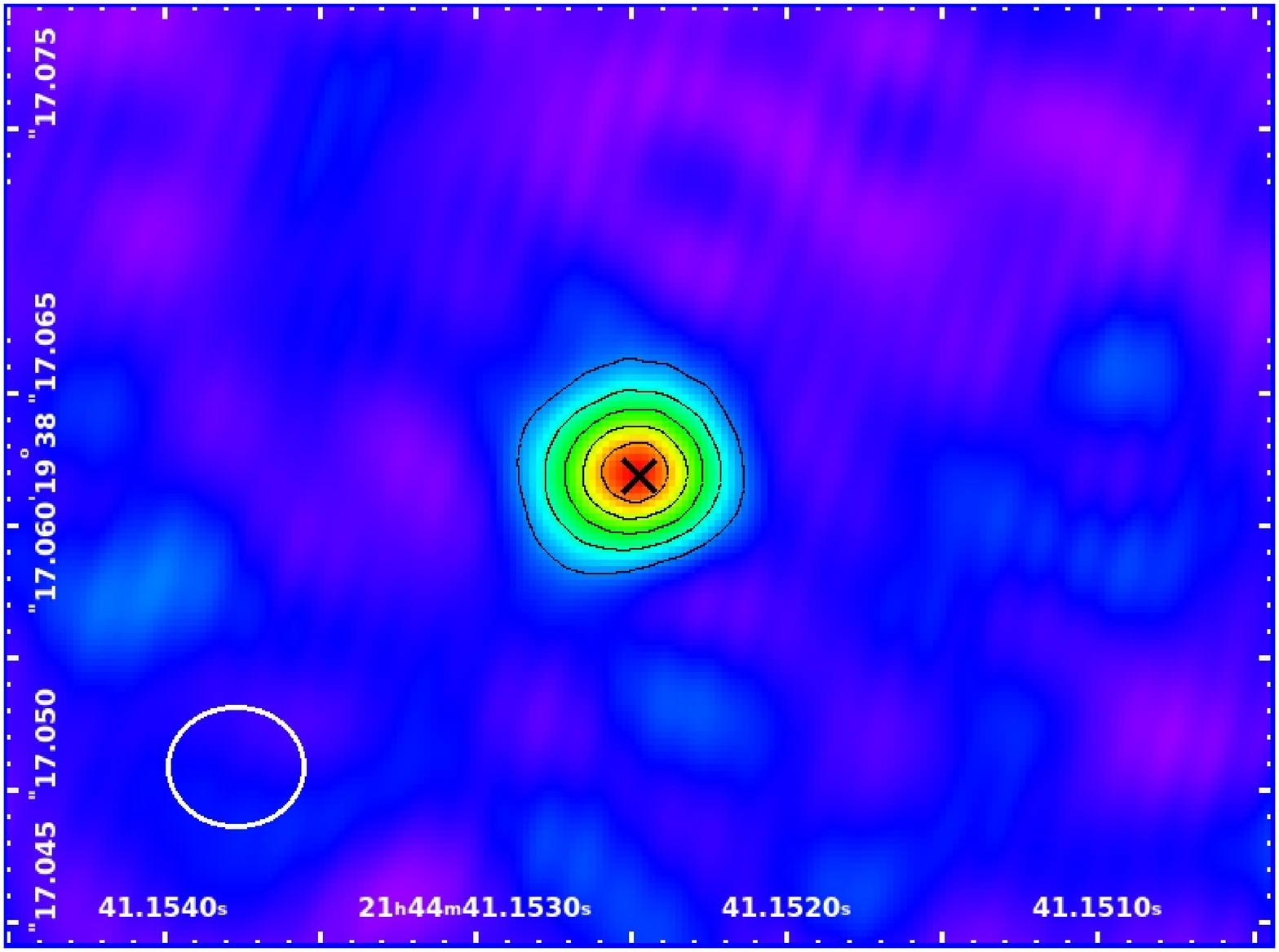} 
\hskip 20 mm
\includegraphics[width=75mm,height=55.7mm]{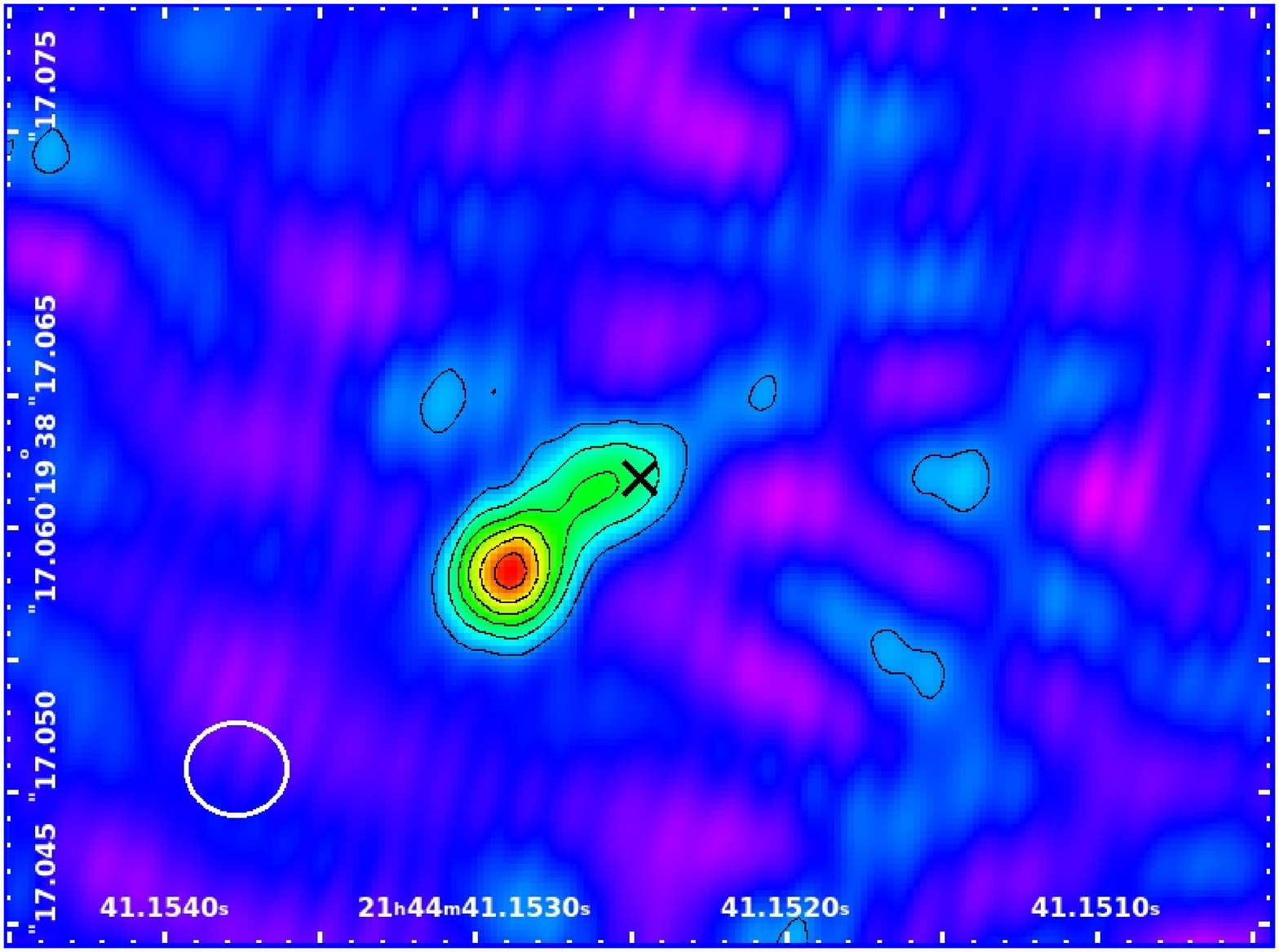} 
\caption{European {\it VLBI} Network observations at 5 GHz: (left) 2013 Feb 22; (right) Feb 23
centred on RA 21 44 41.152497 and Dec +38 19 17.06210. The contours scale linearly with a step of 89.6$\mu $Jy;
the peak flux density/r.m.s. for the epoch 1 and epoch 2 maps are 523/20 and 583/30 $\mu $Jy/beam, respectively.
The cross is marked at the peak of the Day 1 map showing the core peak to have moved on Day 2. Some evidence for a
counter jet may be seen.
}
\label{}
\end{center}
\end{figure*}
data are transferred from the telescopes to the correlator at JIVE, the
Netherlands, on national and international academic communication networks.
The data rate from each telescope was 1024 Mbps, and the link from Jodrell
Bank used a Bandwidth-on-Demand connection from London to JIVE across the
Janet, G\'{E}ANT and SURFnet Networks. This was the first time such a link
has been used for e-VLBI observations and was set up as part of the
NEXPRes project. The EVN telescopes used were the Effelsberg (Germany)
\hbox{100-m,} Jodrell Bank (UK) 76-m, Medicina (Italy) 32-m, Noto (Sicily) 32-m,
Onsala (Sweden) 25-m, Toru\'n (Poland) 32-m, Yebes (Spain) 40-m, Westerbork
(the Netherlands) tied array of 12 x 25-m telescopes and Shanghai (China)
25-m. The frequency band used was 4926.49 MHz to 5054.49 MHz in 8 x 16 MHz
sub-bands in both hands of circular polarisation. 2-bit sampling was used at
the Nyquist rate giving a total data rate of 1024 Mbps. Medicina and
Shanghai both used 1-bit sampling at 512 Mpbs and therefore had the same
frequency coverage. The upper 16 MHz sub-band was dropped at Noto due to link
bandwidth limitations, giving a data rate for that telescope of 896 Mbps.
Initial observations were made each day on BL Lac as a `fringe finder' in
order to find clock offsets, followed by a 5-minute cycle of 2 minutes on
the phase reference source J\th 2134+4050 and 3 minutes on the Cyg\th X-2 target.
Alternate observations of the reference source were dropped by the Jodrell
Bank Lovell 76-m telescope giving a 10-minute cycle period, due to slewing
limitations. The observations on Cyg\th X-2 ran from 09:37 UT to 16:55 UT on
22$^{\rm nd}$ February and 09:22 UT to 17:28 UT on 23$^{\rm rd}$ February, except for
Shanghai where observation terminated at 10:00 and 10:05 UT respectively due
to the source setting.

The data were correlated with the EVN Software Correlator at JIVE (SFXC)
using 8 sub-bands with 4 polarisations, i.e. including cross hands, with 32
frequency channels per band and 2-second integration. The correlated data
were then processed using the automatic pipeline (Reynolds, Paragi and
Garrett 2002) which now uses ParselTongue (Kettenis {\&} Sipior 2012) scripts 
to run AIPS programmes. The data were
initially calibrated using antenna noise temperature and gain tables and
then fringe fitted, to allow frequency dependent slopes across each band and
fringe rates to be corrected. The data from the pipeline were edited to
remove obvious errors and the phase reference source imaged by
self-calibration, and the solutions found for this source used to correct
the phases of the target source. The reference source was unresolved and
strong enough to give good solutions and initial imaging was done using the
Difmap programme in the Cal-Tech VLBI package (Shepard, Pearson \& Taylor
1994). Data were then transferred to AIPS for more detailed calibration,
followed by transfer to Difmap for further imaging and model fitting.
Unfortunately snow storms on 23$^{\rm rd}$ February resulted in data from
Effelsberg and Medicina being lost. The resultant images are shown in
Fig. 1 left and right for the two days, the second image having higher noise
levels as expected. To our surprise and pleasure the second day shows a
double structure suggesting an outburst producing an ejected jet.

The positions and flux densities of the components are given in Table 1.
The 2005 position found by {\it MERLIN} is also shown in the table. The proper
motion from May 2005 to February 2013 is therefore -3.0$\pm 0.68$
mas per year in Right Ascension and -0.64$\pm$ 0.68 mas per year in
Declination. 
\tabcolsep 1 mm                                   
\begin{table}
\begin{center}
\caption{{\it VLBI} Positions of components (J2000); the observation of 2005, May 21 
shown for comparison was made using {\it MERLIN}.} 
\begin{tabular}{llllll}
\hline\hline
Epoch          &Cmpt     &Flux &R.A. &Dec. \\
               &         & (mJy)&(h$\;\;$ m$\;\;$s) &($\degmark \;\;\;$ $^{\prime}\;\;\;$ $^{\prime\prime}$)\\
\noalign{\smallskip\hrule\smallskip}
2013 Feb 22    &single   &0.59  &21 44 41.152497  &38 19 17.06210 \\
2013 Feb 23     &NW       &0.38  &21 44 41.152594  &38 19 17.06168 \\
               &SE       &0.55  &21 44 41.152898  &38 19 17.05822 \\
2005 May 21     &single   &1.7   &21 44 41.154453  &38 19 17.06712 \\
\hline
\end{tabular}\\
\label{tab1}
\end{center}
\end{table}
Absolute position errors are estimated as 0.45 mas on the
22$^{\rm nd}$ and 0.50 mas on the 23$^{\rm rd}$, with relative errors between the two days of
0.3 mas in both coordinates. The position error in the Merlin measurement
was 5.3 mas in each coordinate.
%

The change in structure suggests that a radio jet with a bright head (the SE
component) has been ejected, while emission from the core had weakened by
the second day. The head component moved by 6 mas in position angle 141 deg.
%
\begin{figure}                                            
\begin{center}
\includegraphics[width=68mm,height=86mm,angle=270]{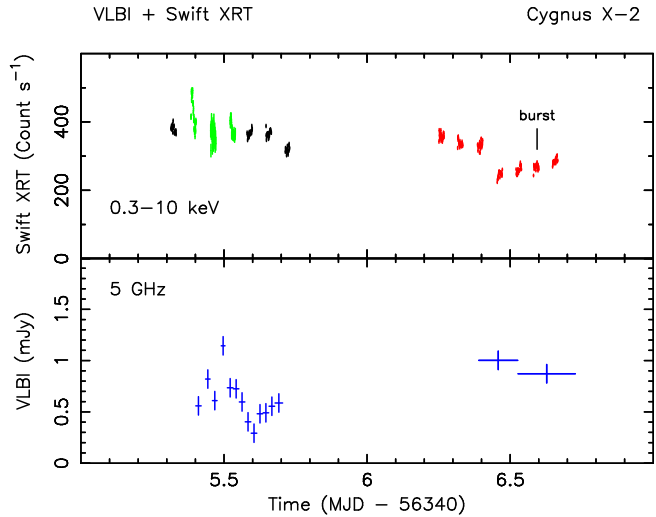}  
\caption{Upper panel: X-ray intensity in {\it Swift} in the band 0.3 -- 10.0 keV;
lower panel: 5 GHz radio fluxes.
}
\end{center}
\end{figure}
A distance for Cygnus\th X-2 was published of 7.2$\pm$1.1 kpc (Orosz \& Kuulkers 1999).
From peak X-ray burst fluxes a distance of 11.6$\pm$0.13  was found by Smale (1998) 
and 11$\pm$2 kpc by Galloway (2008) for an H-fraction X = 0.7. However, not all bursts exhibited the 
radius expansion required for distance determination. We adopt a distance of 9 kpc which gives an
apparent ejection velocity of 0.33$\,$c, taking about 24 hours between the 
measurements.

The signal to noise ratio and aperture plane coverage on the first day were
sufficient to allow model fitting of a circular Gaussian component to the image
for short (0.5 -- 1 hr) sections of data (flux densities shown in Fig. 2). 
The data did not allow accurate determination
of velocity but the position in right ascension suggests that the component 
moved to the east by the end of the run.

The velocity may be further determined as follows.
The loss of the two telescopes on the second day meant that model fitting to
short sections of data was unreliable, however, splitting the data into two
halves showed that the double structure was present in both sections. If we examine
the radio 5 GHz lightcurve (Fig. 2, lower panel), a small radio flare seemed to occur
at around 11:40 UT on Day 1.  Table 2 shows the time of the midpoints of the two sections of
data on Day 2, the flux densities and separation of components in the two halves. If 
we take the mean time for the 1st section of data on the second day as a time at 
which the double structure was clearly established and the time of the weak flare as the 
ejection time then the apparent velocity is v/c = 0.33$\pm$0.12 for a distance of 9 kpc.

\tabcolsep 4 mm
\begin{table}                                           
\begin{center}
\caption{Results of model fitting to 2$^{\rm nd}$ day's data}

\begin{tabular}{llll}
\hline\hline
Date, Time&Component&Flux     &Separation\\
             &         &(mJy)             &(mas)\\
\noalign{\smallskip\hrule\smallskip}
23 Feb 11.11  &core  &0.36  &\\
              &SE  &0.65  &5.016 \\
23 Feb 15.14  &core  &0.43  &\\
              &SE  &0.44  &5.256 \\
\noalign{\smallskip}\hline
\end{tabular}\\
\label{tab2}
\end{center}
\end{table}
The core component has decreased in flux density slightly while the
ejected (SE) component decreases in flux in the second half of Day 2. There is evidence that
the source is still expanding, though at a rate lower than since the
previous day (1.4 mas/day). However, the results show that the source has
expanded by 11:00 UT on 23$^{\rm rd}$ February, 23.3 hrs after the weak flare on
February 22$^{\rm nd}$.

There is some evidence for a counter jet in the radio image (Fig. 1 right). However, 
the peak of the emission to the NW of the source is only at about the 3$\sigma$ level,
and the feature could arise from noise or artifacts. The peak brightness of the clearly seen 
SE jet on Day 2 was found to be $\sim$6 times the peak brightness at the position of the counter 
jet. From this ratio, some constraints may be placed on the jet velocity as discussed in Sect. 4.

\section{The Swift Observations and Results}

Cygnus\th X-2 was observed using the {\it Swift} satellite on February 22$^{\rm nd}$ and 
23$^{\rm rd}$ 2013, as a target of opportunity that allowed simultaneous X-ray and radio 
observations. There were 7 $\sim$1600 s observations spread over 10 hr on the first day, 
beginning at 7:32 UT, then a gap of 12 hr followed by a second similar period of 7 short 
observations. The observations were made in Windowed Timing mode because of the brightness 
of the source. In this mode, the central 200 pixel columns are used while each set of 10 pixel
rows are binned together. The CCDs are then read out with rapid clocking
and the two-dimensional structure is lost, the  
image appearing as a strip of 200 columns only a few units wide. 
On Day 1, in observations 2, 3 and 4 the source was not located optimally, but off the end 
of the strip, as a result of operational constraints. In these cases, a substantial 
number of counts were lost but correction could be made using the point spread function.

The raw data were assembled into an events file using {\sc xrtpipeline}.
Standard screening was applied.
{\it Xselect} was then used to extract a lightcurve from a source region consisting of 
an annulus of inner and outer radii 4 pixel (0.16 arcmin) and 25 pixel (1 arcmin). 
The central part of the source region was not used because of the unacceptable
degree of pulse pile-up in that region. Lightcurves were obtained in a total band
of 0.3 -- 10 keV and also sub-bands of this including 2.5 -- 4.5 keV and 4.5 -- 10.0 keV
to allow construction of a hardness ratio. Background was obtained from a region outside
the source annulus. {\sc xrtlccorr} was used to correct the lightcurves taking into
account vignetting, the point spread function allowing for the different position
angles in different sub-observations. 
This was done for all observations including the 3 observations not optimally directed; 
however, because of the larger corrections made, the systematic errors in these cases were higher. 

The total background-subtracted lightcurve in the band 0.3 -- 10 keV of the whole 2 days 
of observation is shown in Fig. 2 (upper panel), together with the radio 5 GHz data.
Day 1 data are shown in black; sub-observations 2, 3 and 4 are shown in green and 
can be seen with larger errors corresponding to the observations not optimally directed.
Day 2 observations are shown in red.
The X-ray luminosity in the band 1 -- 30 keV was typical for the source,  
varying between 1.3 -- 1.7$\times 10^{38}$ erg s$^{-1}$ during Day 2.
During the second half of the observations, an X-ray burst was seen at the time marked
on the figure lasting a few seconds (and so not seen at 64 s binning).
X-ray bursts are unusual in Z-track sources which have luminosities
more than 10 times higher than those of X-ray burst sources. Type 1 bursts have been seen 
in Cyg\th X-2 over a number of years (Kahn \& Grindlay 1984; Smale 1998; Galloway et al. 2008).


A hardness ratio was formed from the ratio of the counts in the band 4.5 -- 10.0 keV to that
in the band 2.5 -- 4.5 keV and is shown in Fig. 3 as a function of the intensity in the
band 0.3 -- 10 keV. The approximate inferred positions of the normal and flaring branches
are indicated by dashed lines. 
It can be seen that there is no change in hardness ratio while the
intensity changes by a factor of two, showing the source to be on the
horizontal branch. The factor of two is typical and implies that the Hard Apex
between HB and NB will be approximately as indicated by the intersection of the dashed line
with the data. We have re-analysed previous {\it RXTE} data on Cygnus\th X-2 using
the same energy bands as in the present work. The hardness-intensity plot may have a
slightly different shape from that in {\it RXTE} but the horizontal branch remains horizontal
and there is no possibility of confusion with the NB or FB. It is clear that the source 
is on the horizontal branch. This is supported by spectral fitting, a detailed account
of which will be published separately elsewhere.

\subsection{Radio and X-ray correlations}

On the first day, Cyg\th X-2 is already located on the HB as shown by the hardness ratio.
During the observations it moves on this branch, increasing in intensity during the second
sub-observation, moving towards the hard apex of the Z-track. After 2 hours
the radio flare appears. It seems likely that the radio flare corresponds to jet launching.
As expected from this, the integrated radio image for Day 1 does not show a jet. 

On Day 2, there is a substantial drop in X-ray intensity. If the radio data are not integrated
as two halves but left as 7 sub-observations, the first 3 points
indicate a decrease in radio flux, but should be treated with some caution because of
the loss of data from Effelsberg and Medicina. However, these data represent the first
observation of the moment of jet launch in both radio and X-ray.

\begin{figure}                                           
\begin{center}
\includegraphics[width=55mm,height=60mm,angle=270]{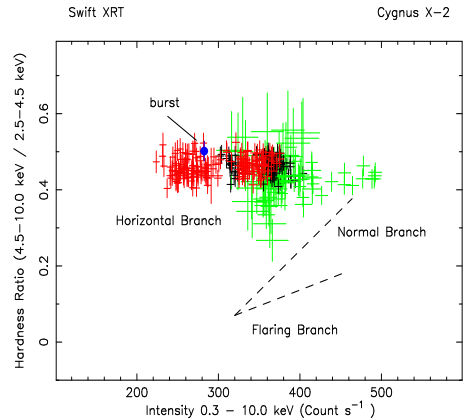}  
\caption{Variation of hardness ratio with intensity. The colours correspond to those in Fig. 2; 
points with larger error bars are observation 2, 3 and 4 from Day 1 (see text).
}
\end{center}
\end{figure}

\section{Discussion}

We have observed the formation of relativistic jets in Cyg\th X-2 in radio and X-rays
and can probably locate the time of jet detachment from the neutron star system at
the radio flare observed on Day 1. Based on this assumption, we find a preliminary value of the 
jet velocity v/c of 0.33$\pm$0.12. This is based on assuming jet formation at the EVN flare
and position of the head at the mid-point of the first radio measurement on Day 2. There is an
uncertainty in the angle moved of 0.4 in 6 mas. We use the maximum possible range of distance
values implied by uncertainties in the quoted values (Sect. 2) giving the above quoted 
uncertainty in v/c of 38\%.

The possible weak detection of a counter jet allows us to place constraints on the jet
for particular assumptions. The ratio of the brightness of the approaching jet $S_{\rm a}$ 
to that of the receeding jet $S_{\rm b}$ when Doppler boosting takes place is given by

\vskip - 2mm
 $$ {S_{\rm a}\over S_{\rm b}} = \left({1 + \beta\, cos\theta\over 1- \beta\, cos\theta}\right)^{k-\alpha}$$

\noindent
where $\beta$ is v/c, $\theta$ is the inclination angle, $\alpha$ is the spectral index and $k$ is a factor 
related to the geometry of the ejecta. We assume $\alpha$ = 0 and $k$ = 3 appropriate to ejection
of discrete condensations. Using the measured ratio of $\sim$6
%
%
and assuming an inclination of 62.5 degrees
(from the UVB light curves, Orosz and Kuulkers 1999) gives an intrinsic 
velocity for the jet of $\sim$0.6$\,$c and an apparent velocity of the approaching jet of $\sim$0.8$\,$c, 
higher than that observed. Possibly
the ejection could be one sided, have unequal velocities for the approaching 
and receding jet or more likely have intrinsic differences in radio flux between the two sides 
- perhaps due to significant absorption. High resolution observations at two or more radio frequencies
would be needed on future ejecta to investigate this possibility.

In the case of Sco\th X-1 there have been extensive investigations of the jets by Fomalont
and co-workers (e.g. Fomalont et al. 2001) and there appear to be distinct similarities as
might be expected as Sco\th X-1 is also a Z-track source although more luminous, having a luminosity
typically three times higher than Cyg\th X-2. Sco\th X-1 exhibits a radio core, and northeast and 
southwest jets with measured average velocities of 0.45$\,$c. We suspect however, that
formation of the relativistic jet is due to heating of the neutron star and the corresponding 
increased radiation pressure rather than other mechanisms such as disk instability.
As the inner accretion disk in the Z-track sources is a thick disk, inflated to a height of
the order of 50 km by the disk's radiation pressure, collimation of the jet is possible
at the conical opening in the hot inner disk, the radiation pressure having components both
vertical and horizontal, the latter having a collimating effect.

\subsection{A model for the Z-track sources}

In the standard approach to the Z-track sources (Hasinger et al. 1990) it is thought that the physical
changes between the branches are caused in some way by changes in the mass accretion rate $\dot M$
which was thought to increase monotonically in the sense HB $\rightarrow$
NB $\rightarrow$ FB. However, the X-ray intensity decreases along the NB
which is not expected if $\dot M$ is increasing.

A new explanation of the nature of the Z-track sources based on extensive analysis
of {\it Rossi-XTE} data was proposed for the Cygnus\th X-2 like group (Church et
al. (2006); Ba\l uci\'nska-Church et al. 2010) and then extended to the Scorpius\th X-1
like group (Church et al. 2012). In this model, the mass accretion rate $\dot M$ does
not increase monotonically round the Z-track as previously postulated, but increases
from the soft to hard apex. The increase
of $\dot M$ heats the neutron star, the blackbody temperature $kT$ increasing from
$\sim$1 keV to $\sim$2 keV. There is a corresponding increase in radiation pressure by
a large factor the neutron star emission becoming super-Eddington which will disrupt 
the inner disk forcing material upwards perpendicular to the accretion disk, so launching 
the jet. Evidence was presented showing that the flaring branch consists of unstable nuclear
burning on the surface of the neutron star.

\subsection{Prediction of jet launching}

In a separate study (Ba\l uci\'nska-Church et al. {\it in prep.}) 15 years of data from the 
{\it RXTE} All Sky Monitor were examined revealing longterm variations of the source very clearly
and significantly on a 
timescale of 40 -- 70 days which were not strictly periodic. 
Wijnands et al. (1996) found a periodic 78-day effect but based on much less data.
Pointed {\it RXTE} observations 
in which the position on the Z-track was known were superimposed on the 15-year lightcurve. It
was found that the Z-track movement only took place when the source was in a high state 
occurring quasi-periodically up to ten times per year. The HB in which jets predominantly 
occur is only seen in these high states. Radio detections of jets were then added, showing 
that these occurred when the ASM intensity increased above 40 count s$^{-1}$. 
We thus now have the ability to predict the launching of jets based on observation of the source 
moving from a low intensity to above a critical value. Since the demise of {\it RXTE}, 
{\it MAXI} can be used in the same way. 
In the present observation the jet was formed when the source had crossed the critical intensity 
giving strong support to our ability to predict jets.

\section{Conclusions}

We have captured jet formation in Cygnus\th X-2 in radio and X-rays for the first time.
The {\it Swift} observations confirm that the source is on the horizontal branch of the
Z-track.

The observations were found to take place at a period of increased X-ray intensity as
monitored by {\it MAXI} supporting our findings from a study with the {\it ASM} and {\it MAXI}
that a minimum critical intensity is needed for jet formation. We believe
that we can now predict jet formation and it should be
possible to trigger observations to observe jet formation in more detail, without the
major uncertainty previously attached to observations at arbitrary epochs.

\section*{Acknowledgments}

We thank Neil Gehrels and the {\it Swift} team for agreeing to the {\it TOO} and Richard
Porcas for his efforts in scheduling the radio observations. We thank 
Matteo Perri and Milvia Capalbi for their help with non-standard 
issues of XRT analysis. The e-VLBI research infrastructure in Europe is supported by the 
EU's 7$^{\rm th}$ Framework Programme under grant 
RI-261525 NEXPRes. The EVN is a joint facility of European, Chinese, South African and other 
radio astronomy institutes funded by the national research councils.

\label{lastpage}


\begin{thebibliography}{}

\bibitem[]{}
Ba\l uci\'nska-Church M., Gibiec A., Jackson N. K., Church M. J., 2010, A\&A, 512, A9

\bibitem[]{}
Church M. J., Ba\l uci\'nska-Church M., 2004, MNRAS, 348, 955

\bibitem[]{}
Church M. J., Halai G. S., \& Ba\l uci\'nska-Church M., 2006, A\&A, 460, 233 

\bibitem[]{}
Church M. J., Gibiec A., Ba\l uci\'nska-Church M., Jackson N. K., 2012, A\&A, 546, 35

\bibitem[]{}
Fomalont E. B., Geldzahler B. J., Bradshaw C. F., 2001, ApJ, 558, 283


\bibitem[]{}
Galloway D. K., Muno M. P., Hartman J. M., Psaltis D., Chakrabarty D., 2008, ApJS, 179, 360

\bibitem[]{}
Hasinger, G., \& van der Klis, M., 1989, A\&A, 225, 79


\bibitem[]{}
Hasinger G., van der Klis M., Ebisawa K., Dotani T., Mitsuda K., 1990,
A\&A, 235, 131

\bibitem[]{}
Kahn S. M., Grindlay J. E., 1984, ApJ, 281, 826


\bibitem[]{}
Kettenis M., Sipior M., 2012, Astrophysics Source Code Library, 
record ascl: 1208.020 


\bibitem[]{}
Migliari S., Miller-Jones J. C. A., Russell D. M., 2011, MNRAS, 415, 2407



\bibitem[]{}
Orosz J. A., Kuulkers E., 1999, MNRAS, 305, 132 

\bibitem[]{}
Penninx, W. 1989, in J. Hunt and B. Battrick, eds. ``Proceedings of the 23rd ESLAB symposium
on Two Topics in X-ray Astronomy'', Bologna, Sept. 1989, 
ESA SP-296, 185


\bibitem[]{}
Reynolds C., Paragi Z., Garrett M., 2002, 
preprint (

\bibitem[]{}
Rushton A. P., Spencer R. E., Strong M., et al., 2007, MNRAS, 374, L47 

\bibitem[]{}
Schulz N. S., Hasinger G., Tr\"umper, J., 1989, A\&A, 225, 48

\bibitem[]{}
Shepherd M. C., Pearson T. J., Taylor G. B., 1994, BAAS, 26, 987

\bibitem[]{}
Smale A. P., 1998, ApJ, 498, L141


\bibitem[]{}
Wijnands R. A. D., Kuulkers E., Smale A. P., 1996, ApJ, 473, L45



\end{thebibliography}
\end{document}